# MoCoNet: Motion Correction in 3D MPRAGE images using a Convolutional Neural Network approach

Kamlesh Pawar [1,2*], Zhaolin Chen [1,3], N. Jon Shah [1,4], Gary F. Egan [1,2]

1. Monash Biomedical Imaging, Monash University, Melbourne, Australia
2. School of Psychological Sciences, Monash University, Melbourne, Australia
3. Department of Electrical and Computer System Engineering, Monash University, Melbourne, Australia
4. Institute of Medicine, Research Centre Juelich, Juelich, Germany

**Running Head:**
MocoNet: Motion Correction in 3D MPRAGE using Convolutional Neural Network

* Corresponding author:

| | |
|---:|---|
| **Name** | Kamlesh Pawar |
| **Department** | Monash Biomedical Imaging |
| **Institute** | Monash University |
| **Address** | 770 Blackburn Road, Notting Hill, VIC-3168, Australia |
| **E-mail** | kamlesh.pawar@monash.edu |

**Manuscript word count:** 3100
**Abstract word count:** 250

# Abstract

**Purpose:** The suppression of motion artefacts from MR images is a challenging task. The purpose of this paper is to develop a standalone novel technique to suppress motion artefacts from MR images using a data-driven deep learning approach.

**Methods:** A deep learning convolutional neural network (CNN) was developed to remove motion artefacts in brain MR images. A CNN was trained on simulated motion corrupted images to identify and suppress artefacts due to the motion. The network was an encoder-decoder CNN architecture where the encoder decomposed the motion corrupted images into a set of feature maps. The feature maps were then combined by the decoder network to generate a motion-corrected image. The network was tested on an unseen simulated dataset and an experimental, motion corrupted *in vivo* brain dataset.

**Results:** The trained network was able to suppress the motion artefacts in the simulated motion corrupted images, and the mean percentage error in the motion corrected images was 2.69 % with a standard deviation of 0.95 %. The network was able to effectively suppress the motion artefacts from the experimental dataset, demonstrating the generalisation capability of the trained network.

**Conclusion:** A novel and generic motion correction technique has been developed that can suppress motion artefacts from motion corrupted MR images. The proposed technique is a standalone post-processing method that does not interfere with data acquisition or reconstruction parameters, thus making it suitable for a multitude of MR sequences.

# INTRODUCTION

Magnetic Resonance Imaging is an established medical imaging modality with the ability to produce detailed images of different tissue types. However, despite its advantages, it is a slow imaging modality and any patient motion during scanning causes artefacts in the MR images. Since the advent of MRI, the removal of motion artefacts has continued to be an unresolved problem for most clinical applications. In particular, motion artefacts present significant difficulties when scanning pediatric, elderly, claustrophobic and stroke patients. Apart from patient discomfort, motion artefacts also result in financial losses (1) for the imaging sites. The current state-of-the-art motion correction techniques can be broadly classified into two categories: navigator sequence-based approaches and external motion detection devices.

The navigator based approaches involve inserting k-space, or image navigators, into the imaging sequences. The navigators are inserted within the sequence at different time points and are compared with a reference to estimate motion parameters. The parameters can subsequently be used to perform either prospective or retrospective motion correction. The first navigator sequence to estimate in-plane linear motion was proposed by Ehman et. al. (2), and involves acquiring the zeroth phase encode in both the x and y directions. The phase difference between navigators at different time points provides a measure of motion in the x and y directions. However, this technique only compensates for linear motion. An alternative technique, known as PROPELLER (3), is a self-navigation technique in which a number of straight lines called blades are acquired at different angles to cover the whole k-space. The acquisition of blades at various angles results in the acquisition of the centre of k-space multiple times; with the points that are sampled multiple times used as navigators to extract rotation and translation parameters. However, a limitation of the PROPELLER technique is that it can only provide in-plane motion parameters. Multiple implementation of the use of k-space navigators to estimate motion parameters in all six degrees of freedom have been developed (4–9) and as well as image navigators (10–13). All navigator sequences must be purposefully designed, with the acquisition time and SNR to some extent negatively affected as well as constraints imposed on protocol parameters, including TR/TE/TI.

External device-based approaches involve estimation of the motion by an external device, such as an optical camera, that tracks a marker placed on the patient (14–18) or by using field probes that measure the magnetic field perturbation due to motion (19,20). The major limitations of external device techniques are that they involve additional cost, they require calibration, and the use of mounting markers on the patient may cause patient discomfort and potential patient compliance issues, making them impractical for clinical usage.

In order to improve on the aforementioned limitations, in this paper we introduce a novel technique for motion artefact suppression. The proposed technique is a data-driven approach, based on a deep convolutional neural network (21) that learns to identify and suppress motion artefacts from the images retrospectively. This technique neither penalises image acquisition parameters nor puts any constraints on TE/TR/TI. Moreover, it is an inexpensive post-processing method, which can be readily incorporated across all scanners, as well as in standalone medical image visualisation software.

Convolutional neural networks (CNN) have been used in MRI for image reconstruction from undersampled data (21–27) and for MR image analysis (28). A wide variety of CNN architectures exist and can be selected based on the particular application of interest. The basic principle of all CNN is the convolution operation. Convolution operations are widely used in image processing applications, such as image segmentation, edge detection, image de-noising, Gaussian blurring, and median filtering, to name a few. For instance, in order to remove noise from images, a simple averaging kernel for convolution can be used, which is equivalent to low pass filtering. In these applications, the convolution kernels are hand-engineered using prior knowledge of the imaging process, noise statistics, and the nature of the artefacts. However, algorithms for such applications including image de-noising and segmentation are being rapidly replaced by deep learning approaches.

We hypothesised that ringing, blurring and ghosting artefacts caused by motion could be suppressed with convolution operations. A set of convolution kernels to remove motion artefacts cannot be predetermined because of the nature of the motion that causes the artefacts. We leveraged the power of data-driven deep learning methods to determine sets of convolution kernels that could efficiently remove artefacts arising due to the motion. The objectives of this study were to artificially simulate motion corrupted images, and to subsequently use them to train a deep learning network that could learn appropriate convolution kernels to identify and correct 3D MPRAGE images for motion artefacts. Our aim was to then apply the convolution network to the motion corrupted patient studies to remove the patient induced motion artefacts.

## THEORY

### *Model of Motion artefact in MR imaging*

Motion during MR data acquisition manifests itself as artefacts in the reconstructed images. The nature of the artefacts found in structural scans depends on the magnitude of the motion and the time point at which it occurred. Consider a movement (i.e. position change) occurs at the time point *t* during data acquisition (i.e. during the $n^{th}$ phase encoding step). The motion corrupted k-space can be considered as a linear combination of two partial sets of k-spaces (Fig.1 (c), (l)) before and after motion occurrence. The corresponding motion corrupted image (Fig. 1 (m)) can then be modelled by a linear combination of motion free images (before and after change of position) convolved with a sampling kernel (see Fig 1 for an example). Mathematically, the motion corrupted image can be represented as:

$$I_m = \sum_{t=0}^{t=T} I_t \circledast h_t = \sum_{t=0}^{t=T} (M_t I_0) \circledast h_t \qquad (1)$$

where, $I_m$ is the motion corrupted image, T is the number of times the subject moved, $I_t$ is the object position at time point *t*, $I_0$ is the motion-free image, $M_t$ is the motion matrix at time *t*, $\circledast$ is the convolution operation and $h_t$ is the convolution kernel determined by the time point *t*.

Inspired by the similarity between the convolutional motion model presented here and convolutional neural networks, we investigated the use of deep learning CNN to solve the motion correction problem. In order to estimate the motion-free image $I_0$ in Eq. (1), the process of training was applied to estimate a deep learning model that could identify and suppress the motion artefacts.

*Deep Learning Network Architecture*

The encoder-decoder Unet (29) architecture (Fig. 2) was designed for motion correction. The encoder network consists of a series of convolution and pooling layers with three convolutions before every pooling operation. The decoder consists of a series of convolution and up-sampling layers. There are three convolutions before every up-sampling operation. The decoder network consists of dropout layers after every convolution with a dropout fraction of 0.2. The last layer in the network is a mean squared loss layer, which computes the mean squared error between the output and the target image (no motion image). The input to the network is a 256×256 motion corrupted image and the output is the motion corrected image.

**METHODS**

*In vivo* experiments were performed on a Siemens Skyra 3T MRI scanner (Siemens Healthineers, Erlangen, Germany) with a maximum gradient strength of 45 mT/m and a maximum slew rate of 200 mT/m/ms. For experimental validation of the proposed motion correction technique, 3D MPRAGE datasets were acquired with TE/TR = 3/2000 ms, FOV =

256x256x192 mm$^3$, resolution 1 mm isotropic, flip angle = $10^0$. Informed consent was obtained from all volunteers in accordance with the Institution's human research ethics policy.

*Motion Simulation Datasets*

3D MPRAGE datasets without motion were acquired from 44 subjects and used to generate a training dataset, and acquisitions for a further 11 subjects were used to generate test datasets. Five different random motions were simulated for each subject, thus providing 220 sample images for the training cohort and 55 sample images for the test cohort. In order to generate motion corrupted images, 3D k-space datasets of the MPRAGE images were distorted with a simulated 3D rigid body random motion as described below.

Motion simulation was performed with 3D rigid body random motion between ±5 mm of translation and ±5$^0$ of rotation on all the three axes. For each subject the k-space dataset was simulated to move randomly either one or two times during the acquisition. MPRAGE images of size 256×256×256 were used for the simulation purpose and the motion was simulated along the slowest acquisition direction which was the phase encode direction (anterior-posterior). The k-space datasets at the different positions and orientations determined by the simulated motion were calculated. The blocks of k-space for the different simulated positions were gathered to generate a motion corrupted simulated k-space dataset. The motion corrupted simulated k-space dataset for each subject was Fourier transformed to generate motion corrupted image.

For each subject both the motion corrupted image and the motion-free images were normalized to 0.5 mean and 1/6 standard deviation. Additionally, the background was estimated using a thresholding, followed by dilation and erosion operations, and the estimated background was set to zero. After processing through the deep learning network, each output motion corrected image was renormalized to its initial mean and standard deviation. The CNN was trained to suppress motion artefacts from individual 2D axial slices of the 3D MPRAGE dataset. The following parameters were used for training the network in the Keras deep learning library with Tensorflow backend: RMSprop optimiser with initial learning rate = 0.001, rho = 0.9, decay = 0.0, batch size = 4 and a total of 75K iteration on NVIDIA Tesla P100 GPU.

*Motion Corrupted Datasets*

A total of eight scans with real motion were acquired from four subjects with two scans for each subject. In order to acquire motion corrupted data, each subject was instructed to move during the scan. The subjects were asked to perform a small random motion two times during the eight minutes of scanning protocol. The reconstructed images from the motion corrupted k-space datasets were processed by the trained network and motion corrected images were obtained.

*Evaluation Matrices*

Two metrics were used for evaluation of the motion correction in the motion corrupted images: (i) percentage error for simulation where the ground truth was known and (ii) Naturalness Image Quality Evaluator (NIQE) score for the experimental results where the ground truth was not known. In the absence of an absolute ground truth for the motion corrupted datasets, it was not possible to compute the percentage error or similar metrics to evaluate the performance of the motion correction. In order to quantitatively evaluate the quality of the motion correction for the experimental dataset, we compared the motion corrupted images and the motion corrected images with the NIQE score (30), which evaluates the perceptual quality of an image. A lower NIQE score signifies a better perceptual quality of an image. The spatial domain natural scene statics (NSS) model for the NIQE was modified for use with MR images.

**RESULTS**

The magnitude of the motion was observed to determine the severity of the artefacts in the reconstructed image. The time at which the motion occurred during the acquisition also determined the severity and nature of the artefact. For example, motion occurring in the outer regions of the k-space (high-frequency component) manifested as ringing in the reconstructed image (Fig. 3 (a-d)). Motion occurring near to the centre of the k-space (low-frequency component) manifested as blurring (Fig. 3 (e-h)). In Fig. 3 (a-d), the time at which motion occurred remained fixed and the magnitude of the motion was increased. It can be observed that the ringing increased and the spatial location of the ringing changed as the magnitude of the motion increased. In Fig. 3 (e-h), the time point at which motion occurred was varied and the magnitude of the motion remained fixed. It can be observed that the blurring increased as the motion time point moved towards the centre of the k-space. From this simulation, it can be inferred that the time at which the motion occurs largely determines the nature of the motion artefacts (ringing or blurring), and the magnitude of the motion largely determines the spatial distribution of the motion artefacts.

Feature activation maps of the CNN layers marked as d1, d2, d3, and d4 in Fig. 2 are shown in Fig. 4. The activated features in the CNN encoder layer d1 and d2 extracts basic features including edges, the foreground, and the background from the image. The features in the encoder layers (Fig.4 (d1-d2)) consist of the motion artefacts. The activated feature map in the decoder layers (Fig.4 (d3-d4)) demonstrate that the motion artefacts have been removed. The features in the decoder network are formed by combining the features in the encoder network. The encoder network learned to combine the encoder features to reject artefacts and to combine only useful information. Notably, layer d3 learned to segment white matter and grey matter in the image, with these segmented regions subsequently combined to reconstruct artefact free images.

The performance of the trained DL network was evaluated on the unseen test datasets of 55 subjects. For simulated motion of ±5 mm translation and ±5$^0$ rotation the percentage error depended on both the magnitude of the motion (Fig. 5 (a-b)) and the time of the motion (distance from the centre of k-space, Fig. 5 (c)). The overall error varied between 1.5 – 5.5 % (Fig. 5 (d)). When the motion occurred near the centre of k-space the error in the output images tended to be relatively higher. Whereas, when the motion occurred in higher frequency regions of k-space, the output images had comparatively lower percentage error. The overall error (Fig. 5 (d)) can be considered as a combination of the two errors, with one decreased as the time of motion moved towards high frequency in k-space and the second due to random motion.

Examples of the motion corrected images for four test subjects, along with the percentage errors are shown in Fig. 6. In the first three rows, the reconstruction error is less than 2.81 %, where the motion was either small or occurred at a location away from the centre of k-space. In row 4, the motion occurs close to the centre of the k-space, which resulted in prominent artefacts in the input image, and consequently, a larger error in the motion corrected image. It can be deduced from the simulation results (Fig. 5 and Fig. 6) that the time point at which motion occurred played an important role in determining the nature of the artefact, and subsequently, the performance of the proposed deep learning motion correction method.

Experimental validation was performed on datasets acquired using the MPRAGE sequence. The subjects were asked to perform a small random motion either one or two times at a random time point during the scan. The motion-corrupted images were reconstructed by the default reconstruction pipeline of the scanner and were processed by the trained CNN. Visual inspection of the images (Fig. 7) reconstructed from the CNN demonstrate substantially lower artefacts due to motion, compared to the standard reconstruction. Table 1 shows the NIQE score for the images corrected with the proposed motion correction method for the eight scans. The NIQE score for the motion corrected images were consistently lower than for the corresponding motion corrupted images, and the average percentage improvement was 49%, which demonstrates the significant improvement in the image quality of the motion corrected images.

**DISCUSSION**

We have demonstrated mathematically that the motion artefact MR images can be modelled as a linear combination of convolution between transformed images and convolution kernels. This convolution motion model forms the basis of using convolutional neural networks to learn a CNN model that can correct for motion artefacts. Our simulation results verified that motion during the acquisition of data in high spatial frequency regions of k-space results in ringing, whereas motion during acquisition in low spatial frequency k-space regions results in blurring and ghosting. The encoder in the deep learning network separates the motion corrupted images into features consisting of motion artefacts and non-motion corrupted images. The neuron in the decoder network only allows the motion free features to be integrated into the final motion corrected images. When the images are severely corrupted by motion, predominantly due to motion at low-frequency k-space, the trained network produces images with a slightly larger error compared to similar motion at high-frequency k-space. Although the network was trained on simulated motion corrupted images without considering any spin history effects, the network was able to correct for the real motion, signifying the robustness of the DL network.

Current state-of-the-art motion correction methods that rely on the image or k-space navigators interfere with the MR protocol to the detriment of contrast, scan time, and SNR. Other methods based on external sensors need calibration and are expensive and complex to set up. The method presented in this work is an innovative solution to the problem of motion in MRI. First, unlike navigators, it requires neither the development nor modification of acquisition sequences or image reconstruction methods. Second, it does not interfere with MR protocol parameters such as T1/T2/TI/TE, scan time or contrast. Third, it does not require any external hardware devices, thus making it simple and relatively inexpensive to implement.

We have presented a novel method based on a deep convolutional neural network that is capable of removing motion artefacts from motion corrupted MR images. The method was validated on both simulated and experimental datasets. We have demonstrated that a DL network trained on simulated motion images generalised and was able to correct for real motion that occurred during scanning of a cohort of subjects who were asked to move during the MR scan. The method can work independently without the need for any motion parameter acquisition setup. Alternatively, it can also be used as an adjunct method together with existing motion correction methods, including navigator echo sequences and external motion detection devices.

The current implementation has been validated for the T1 MPRAGE sequence. The current technique has a number of limitations. Firstly, the technique only corrects for sudden motion as the network has not been trained to correct for gradual or continuous motion that may occur during an MR acquisition. Secondly, the motion corrupted images were generated by distorting the k-space data based on the simulated motion parameters, but spin history effects and the Bloch equations were not considered in the simulation. However, although the simulations were performed in the digital domain the proposed method performed competently for the experimental scans in which the subjects moved. Future work will involve the development of improvements to the robustness of the proposed method by integrating continuous motion, generating motion corrupted training images based on the Bloch equations and by validating other contrasts including T2/T2*/proton density images.

**CONCLUSION**

An efficient data-driven method based on a deep convolutional neural network has been developed for MR motion correction, without the need for external sensors or internal navigators. In this work, we have applied the method to 3D T1 MPRAGE datasets. The method can be further extended to other image contrasts or multiple image contrasts, and to other motion types such as non-rigid motion.

*Table 1: Naturalness Image Quality Evaluator (NIQE) scores*

| Subject Number | Scan Number | NIQE before Motion Correction | NIQE after Motion Correction | Percentage Improvement NIQE Score (%) |
|---|---|---|---|---|
| 1 | 1 | 8.52 | 4.74 | 44.4 |
| 2 | 1 | 9.67 | 4.85 | 49.8 |
| 3 | 1 | 10.38 | 4.84 | 53.4 |
| 4 | 1 | 10.15 | 5.02 | 50.5 |
| 1 | 2 | 9.55 | 5.14 | 46.2 |
| 2 | 2 | 9.29 | 4.89 | 47.4 |
| 3 | 2 | 11.00 | 5.30 | 51.8 |
| 4 | 2 | 9.88 | 5.24 | 46.9 |
| Mean Values | | 9.81 | 5.00 | 48.80 |

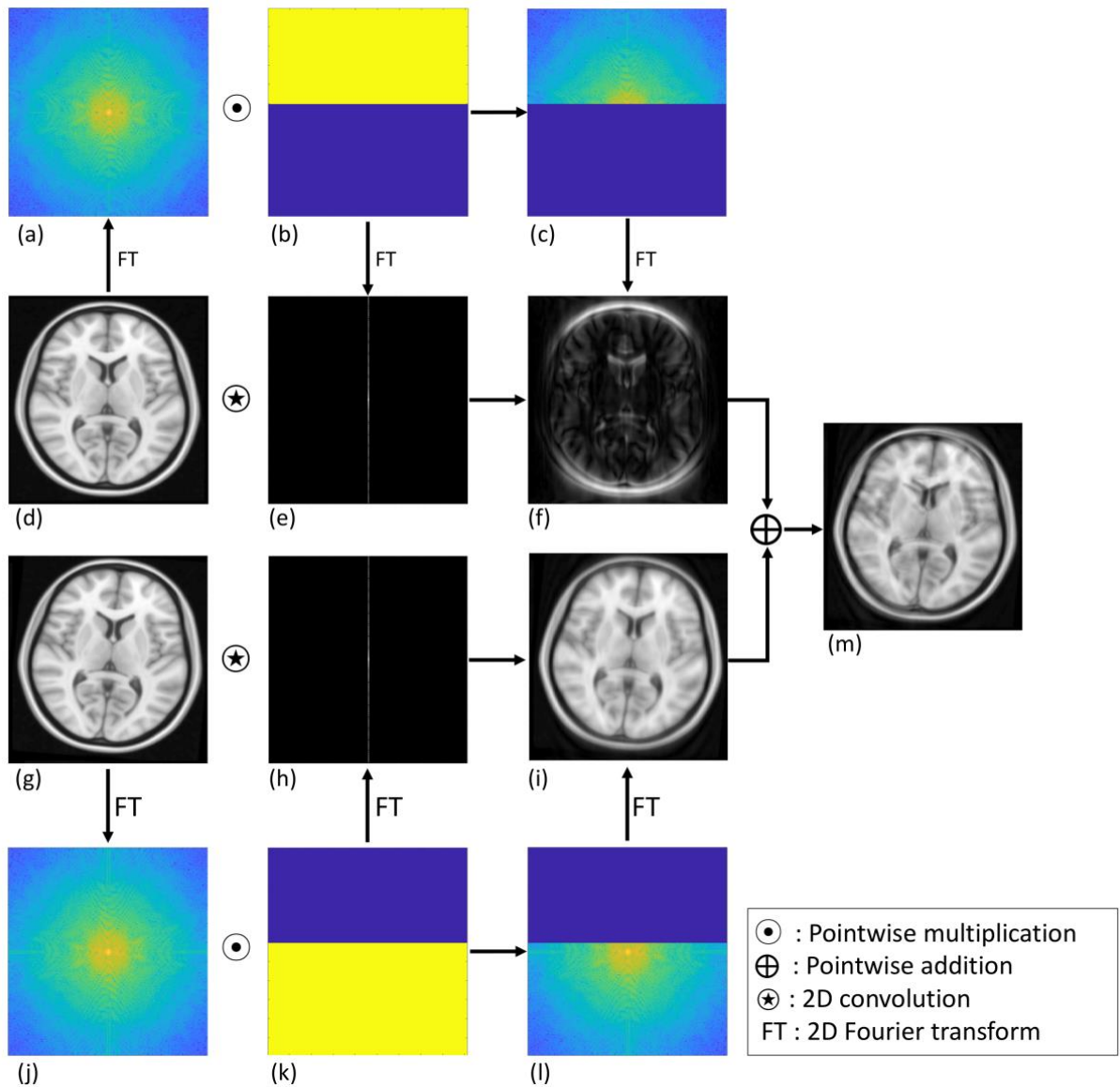

*Figure 1:* Convolution motion model: showing that the motion corrupted image can be modelled as a linear combination of convolution of transformed fully sampled images (d, g) and convolution kernels (e, h). **(a):** k-space of fully sampled image; **(b):** k-space sampling mask (yellow represents 1 and blue represents 0); **(c):** partial k-space of the object; **(d):** image without motion; **(e):** convolution kernel determined by the sampling mask (b); **(f):** image formed by partial k-space (k-space centre is missing); **(g):** image of interest rotated and translated; **(j-l), (h-i)** is same as **(a-c), (e-f)** respectively for the rotated and translated image (g); **(m):** the motion corrupted image.

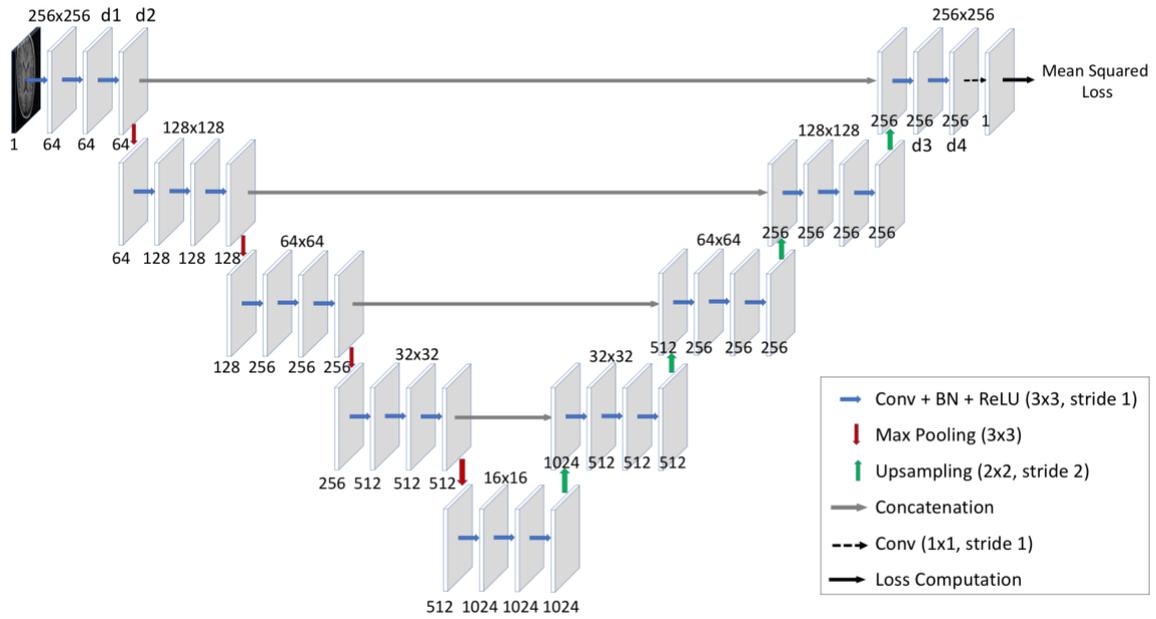

*Figure 2:* *Network Architecture: Encoder-Decoder convolution neural network architecture for motion correction. The input to the network is a 256x256 motion corrupted image. The filters at the top of the figure show the size of the data and the filters at the bottom show the number of filtered outputs. The last layer is the classification layer that predicts the quantized image intensity level for the motion corrected image. The activation of the layers marked as d1, d2, d3, and d4 are shown in Fig. 4.*

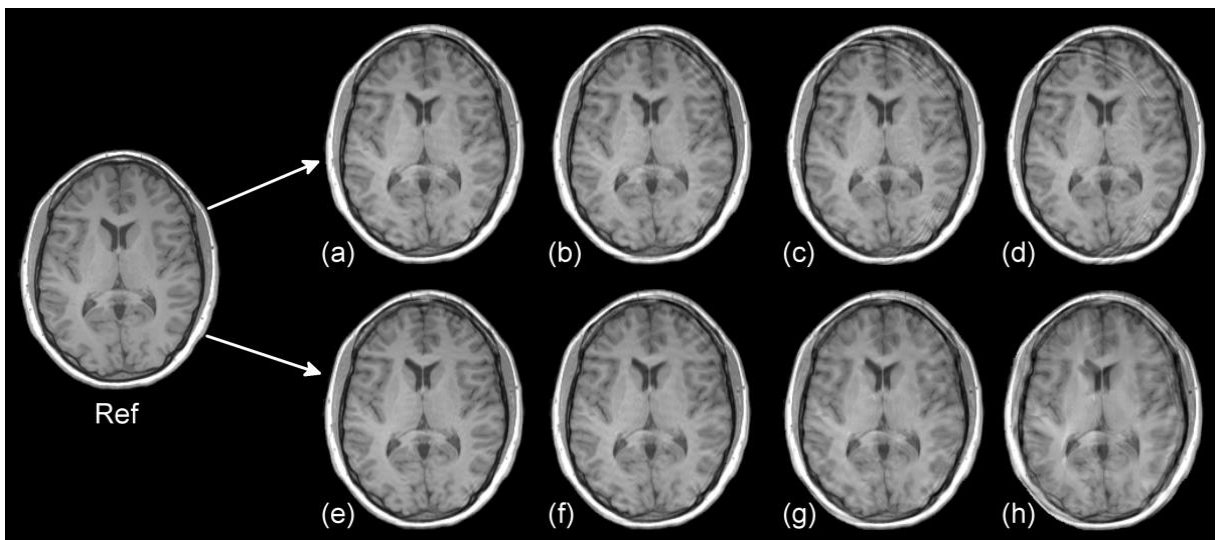

*Figure 3:* *Simulation demonstrating the effect of motion on reconstructed images;* **Ref:** *represents reference image;* **(a), (b), (c)** *and* **(d)** *showing reconstructed images from increasing the magnitude of the motion at a fixed phase encode (100$^{th}$) in the acquisition process, the magnitude were increased as 1, 2, 4, and 6 times respectively;* **(e), (f), (g)** *and* **(h)** *showing reconstructed images due to fixed amount of motion at different phase encoded in the acquisition process, the phase encodes were 90$^{th}$, 100$^{th}$, 110$^{th}$ and 120$^{th}$ respectively.*

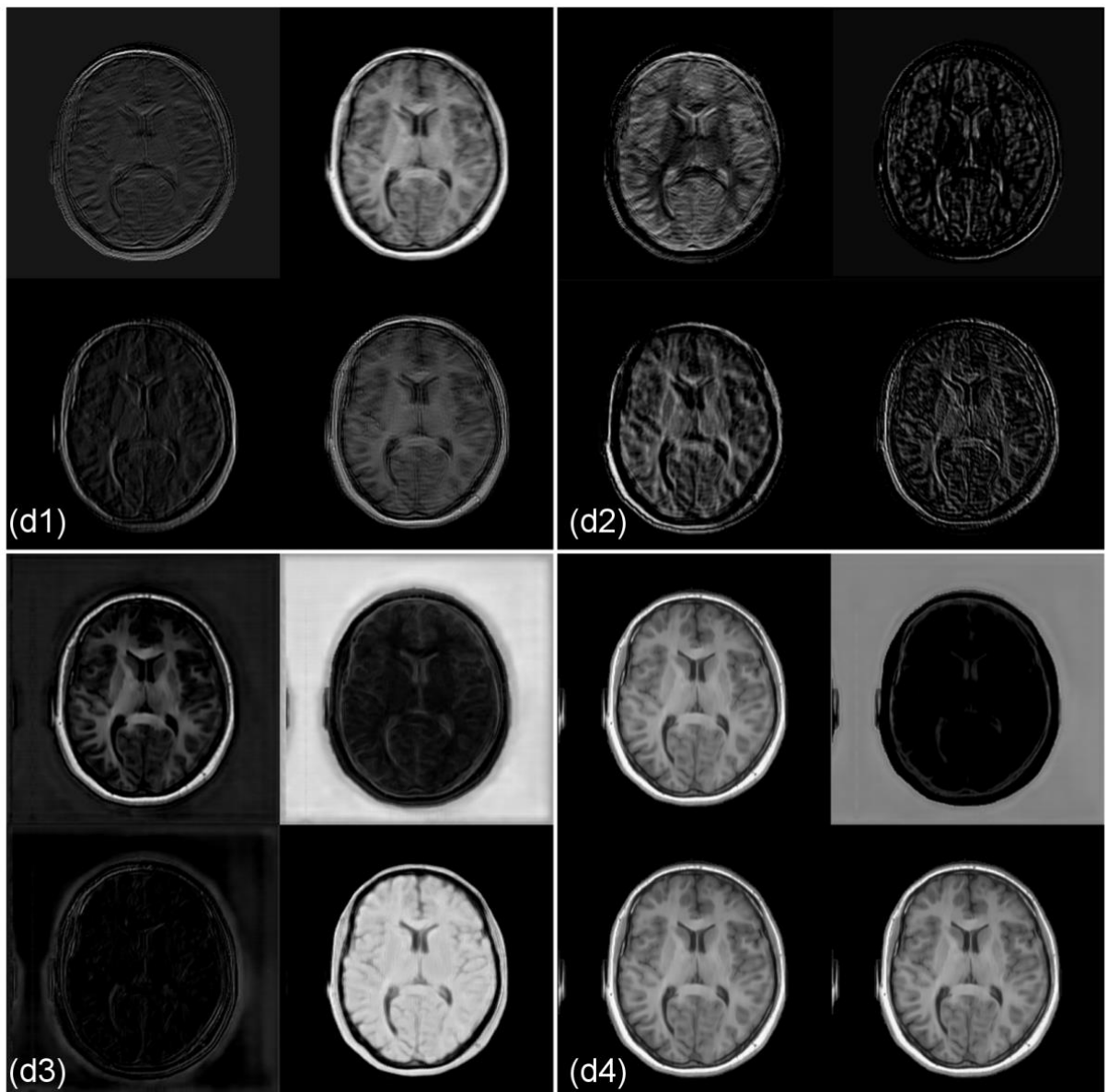

*Figure 4*: Feature maps in the decoding layers d1 and d2, and the encoding layers d3, and d4 of the CNN shown in the Fig. 2. The activated features (bright maps) in the encoding layers d1 and d2 consist of motion artefacts, whereas the activated features in the decoder layers d3 and d4 are not corrupted with motion artefacts.

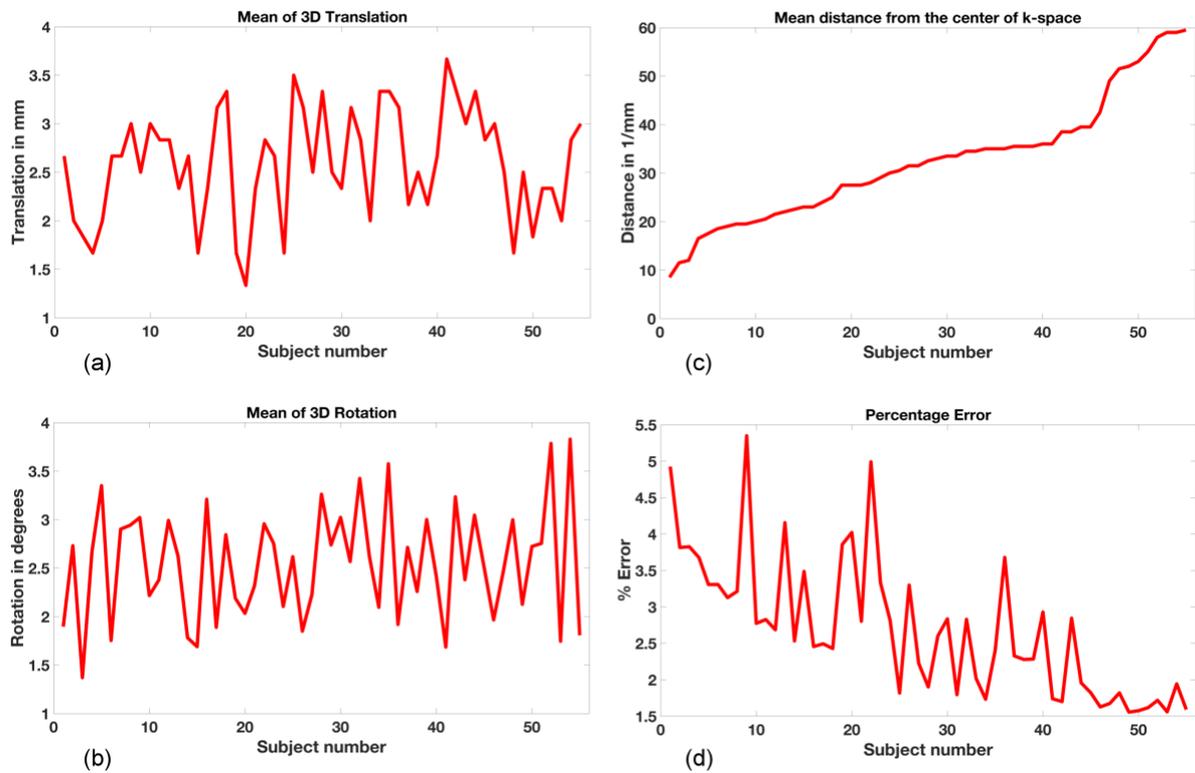

*Figure 5:* Quantitative analysis of the motion suppression performance; **(a):** the absolute mean of the three translations along (x, y, z) axes; **(b):** the absolute mean of all three rotations along (x, y, z) axes; **(c):** the mean distance from the centre of the k-space where the motion occurred; and **(d):** percentage error between the motion corrected and no motion images. For these plots subjects were reordered according to the mean distance from the centre of k-space.

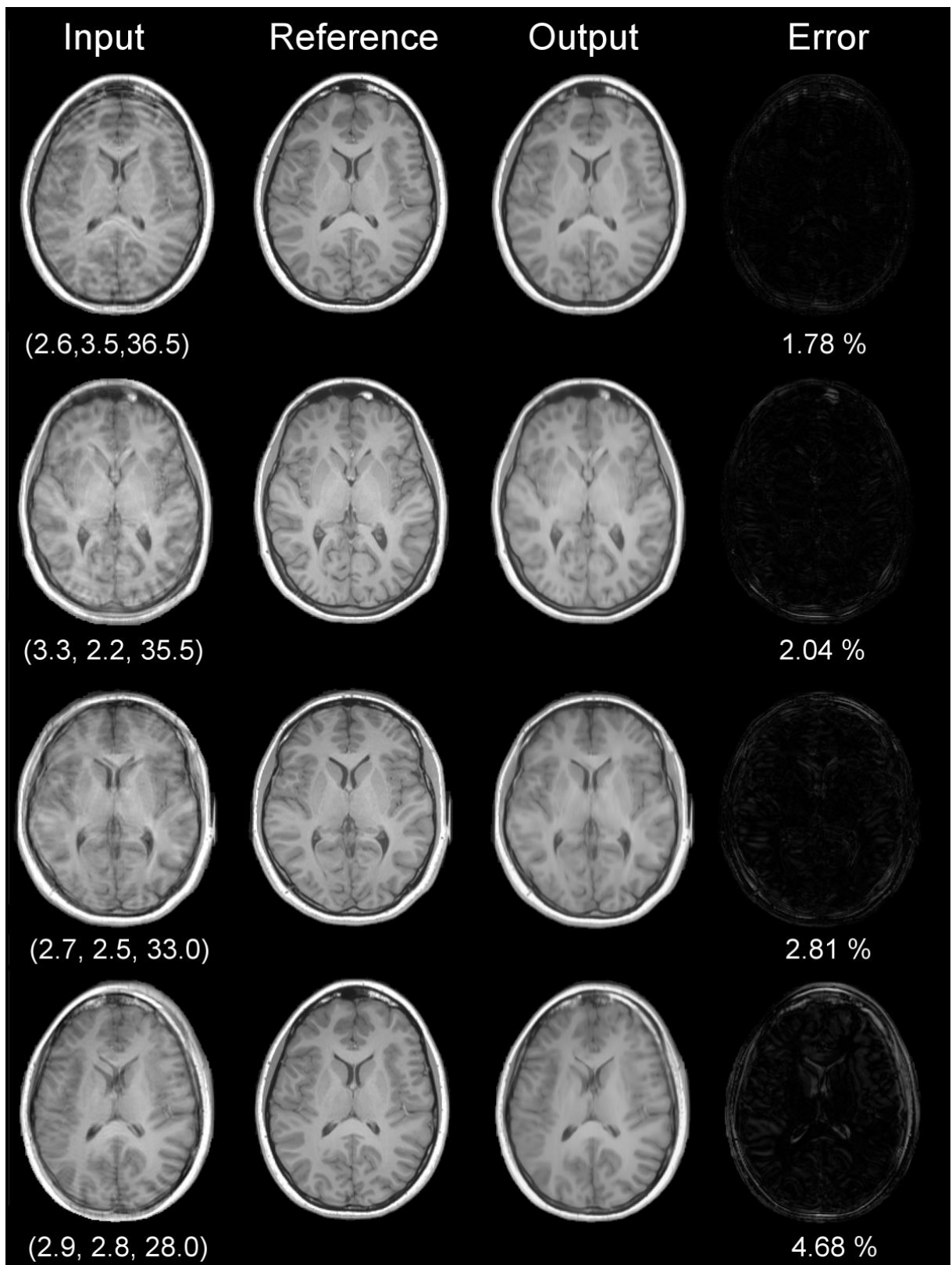

*Figure 6: Motion correction results in a simulated motion dataset; **first column:** Motion corrupted images with (mean rotation, mean translation, mean distance from k-space centre); **second column:** reference images with no motion; **third column:** Motion corrected images after processing through the trained network; **fourth column:** Error images between motion corrected and motion-free images.*

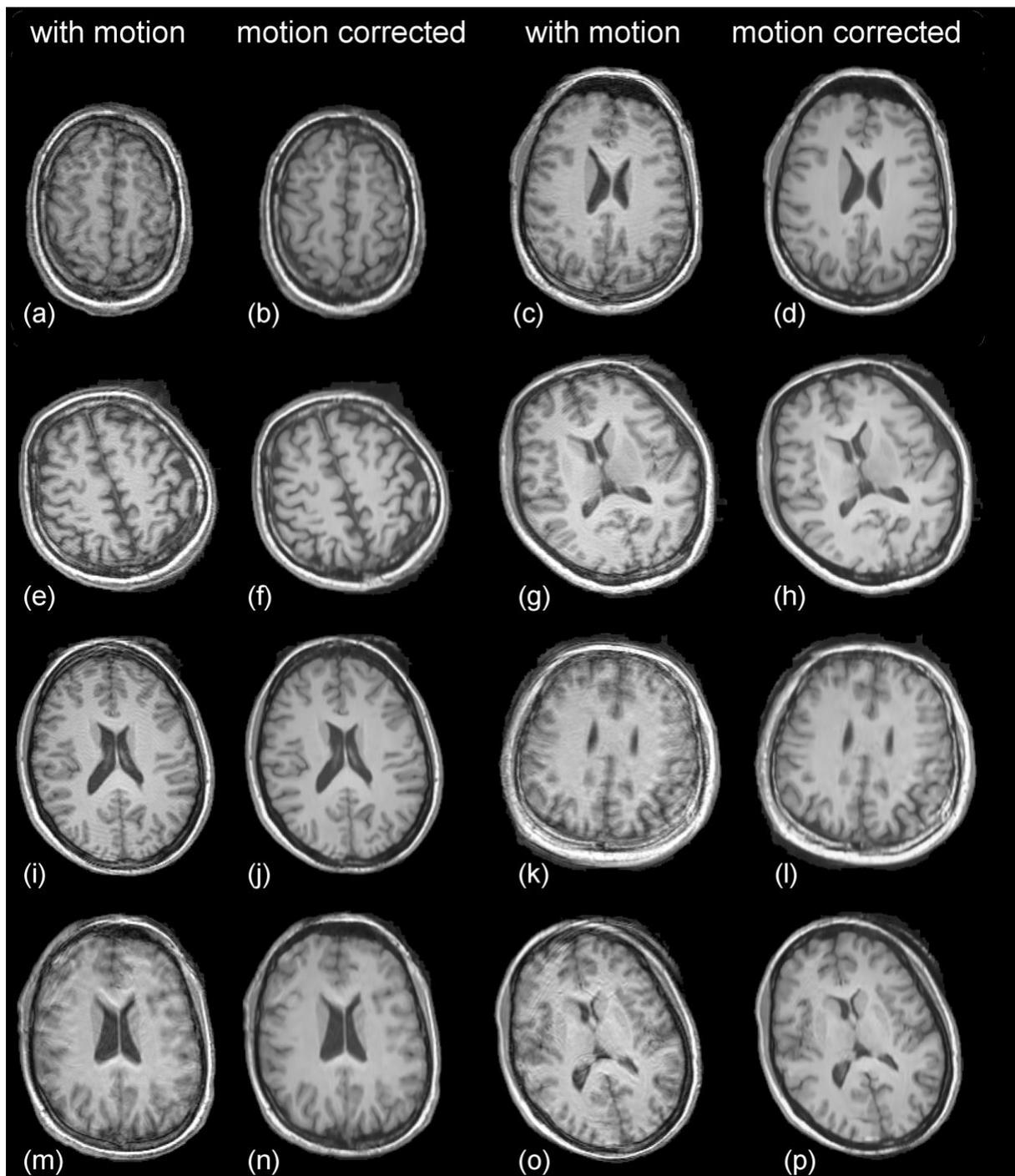

*Figure 7*: Motion correction results in an experimental motion dataset; the subjects were instructed to move two times during the scan at random time points.